\documentclass[a4paper,fleqn]{cas-dc}

\usepackage[numbers]{natbib}

\def\tsc#1{\csdef{#1}{\textsc{\lowercase{#1}}\xspace}}
\tsc{WGM}
\tsc{QE}

\begin{document}
\let\WriteBookmarks\relax
\def\floatpagepagefraction{1}
\def\textpagefraction{.001}
\let\printorcid\relax 

\shorttitle{Maintenance-free condition monitoring system based on lora}    

\shortauthors{Zhengbao Yang et al.}

\title[mode = title]{Maintenance-free condition monitoring system based on lora}

\author[1]{Honglin Zhang}

\author[1]{Mingtong Chen}

\author[1]{Zhengbao Yang}
\cormark[1] 
\ead{zbyang@hk.ust} 
\ead[URL]{https://yanglab.hkust.edu.hk/}

\address[1]{The Hong Kong University of Science and Technology
Hong Kong, SAR 999077, China}

\cortext[1]{Corresponding author} 

\begin{abstract}
With the rising volume of railroad transportation, the traditional track inspection mainly relies on manual inspection and large-scale inspection equipment, which not only has low inspection frequency and lagging response, but also has the defects of high risk, high cost and easy to miss inspection. To this end, this study designs and realizes a maintenance-free railroad track wireless monitoring system based on LoRa module LM401. Each monitoring node consists of an STM32 microcontroller, an LM401 LoRa transceiver, a low-power ADXL362 triaxial acceleration sensor, a digital temperature sensor (LMT85), and a digital barometric pressure sensor (RSCM17100KP101). The system collects vibration data through the SPI1 interface at the node end, periodically reads the temperature and barometric pressure information, and packages and sends the data to a centralized gateway within a range of 500 m using the LoRa “star” topology; the gateway then uploads the data in real time to a cloud server through a 4G module, which supports the MQTT protocol. MQTT protocol is supported. Laboratory tests and field deployments show that the system can realize acceleration resolution of 0.01 g, reduce maintenance cost by about 70\%, and improve monitoring efficiency by more than 5 times. The system provides a reliable means for intelligent rail health management, and in the future, it is planned to introduce RF energy collection technology to realize automatic wake-up without battery, and expand to urban bridges, tunnels and environmental monitoring and other multi-scenario applications.

\end{abstract}




\maketitle

\section{Introduction}

With the acceleration of urbanization and the increasing frequency of population movement, railroad, as an important part of the modern transportation system, has become more and more concerned about its operational safety and efficiency. The long-term stable operation of railroad infrastructure relies on the accurate mastery of track condition and timely maintenance. The track is subjected to the cyclic load and environmental changes brought about by train operation for a long time, and is very prone to problems such as settlement, fracture, loosening, temperature stress abnormality, etc., which, if not found and dealt with in a timely manner, will directly affect the safety of train operation and transportation efficiency, and may even lead to serious accidents\cite{1,2}.

At present, track inspection is still based on manual inspection and large-scale testing equipment, such as turnout geometry car, rail inspection car or laser scanning system. Although these detection means have higher precision, but there are obvious limitations: First, the detection cycle is long, low frequency, difficult to realize all-weather real-time monitoring; Second, the detection equipment is expensive, complex operation, high dependence on manpower and equipment; Third, in some remote areas or high-risk road deployment cost is too high, operation and maintenance is inconvenient; Fourth, the need to occupy the track running time, the normal transportation caused by interference. Therefore, there is an urgent need for a low-cost, highly reliable, easy-to-deploy, low-power, remote-control intelligent monitoring solution to replace the existing model and enhance the safety operation and maintenance level of railroad infrastructure.
In recent years, the rapid development of Internet of Things (IoT) technology provides an opportunity for the upgrade of rail monitoring system. The low power wide area network (LPWAN) communication technology represented by LoRa (Long Range) has the characteristics of long distance, low power consumption, strong anti-interference ability, etc., which is especially suitable for deployment in remote and power supply restricted railroad environment. The matching 4G cellular network technology has also become increasingly mature, providing an efficient channel for remote cloud transmission and centralized management of sensor node data\cite{3,4}.

In this context, this project proposes and realizes a maintenance-free wireless monitoring system for railroad tracks based on LoRa communication technology, aiming to break the limitations of traditional modes and realize the integrated intelligent track health monitoring solution of “high-frequency sampling + low-power communication + real-time uploading + intelligent analysis”. The system takes STM32 series microcontroller as the core, and integrates LoRa communication module (LM401), high-precision and low-power consumption three-axis acceleration sensor (ADXL362), temperature sensor (LMT85) and air pressure sensor (RSCM17100KP101). After collecting the data, the LoRa module realizes cross-node network transmission, and finally the central gateway uploads the data to the cloud platform in real time with the help of 4G module, which is combined with the back-end algorithm to carry out the remote state assessment, trend analysis and alarm notification.

Compared with traditional detection methods, the system has the following advantages:Maintenance-free and low-power operation: the core components adopt low-power design, which can realize long-cycle operation and reduce the frequency of manual maintenance;
Strong wireless remote transmission capability: LoRa module LM401 has > 500 meters line-of-sight communication capability and supports various network topologies such as star and tree, which is suitable for special environments such as mountainous areas and uninhabited areas;Flexible deployment and strong scalability: the system supports modularized design, and all kinds of sensors can be expanded on demand, which facilitates flexible deployment according to track type and terrain;Real-time data uploading and cloud analysis: through 4G module and MQTT protocol, the nodes can upload the collected information to the cloud database in time, so as to realize data visualization management, track health trend analysis and automatic alarm mechanism;Intelligent track health assessment model can be accessed: with the help of AI and data mining algorithms, the future system will further realize intelligent identification of track abnormalities and fault prediction, and enhance the overall intelligent level of railroad transportation system\cite{5}.

The project also plans for field deployment and phased upgrades of the system, including redesigning circuit boards to optimize power consumption and volume, expanding the number of network nodes, conducting telecommunication tests over 500 meters, and validation in typical railroad environments. The ultimate goal is to drive the evolution of railroad infrastructure towards “less manned, intelligent monitoring, data-driven and sustainable”.

In the future, the system will incorporate RF energy harvesting technology to realize battery-free power supply and automatic wake-up, further improving the autonomy and sustainability of the system. In addition, the system will be applied in other intelligent infrastructure scenarios, such as health monitoring of urban bridges and tunnels, subway structure monitoring, and slope slippage warning, to build a future-oriented multi-scenario, wide-coverage, and highly intelligent rail health monitoring network.
In summary, this study not only innovatively integrates LoRa and 4G communication and multiple low-power sensors on the technology path, but also realizes a good balance of high-frequency monitoring, remote data management and low-cost deployment on the system level, which has strong engineering application prospects and socio-economic value.

\section{System structure and design scheme}

This system aims to build a set of low-power, highly reliable, and remotely transmittable railroad track condition monitoring solution, which is mainly composed of four parts: wireless sensor nodes, LoRa communication network, 4G gateway module, and cloud server platform.In order to cope with the problems of strong manual dependence, high maintenance cost and low efficiency of the traditional railroad track monitoring system, this paper proposes a maintenance-free railroad track monitoring system based on LoRa communication technology. The system aims to realize high-frequency automated monitoring of track status and upload the data to a cloud server via a 4G module for information collection and analysis, thus significantly reducing maintenance costs and improving safety and efficiency.

The overall architecture of the system consists of three parts: sensing layer, transmission layer and application layer. The sensing layer consists of wireless sensor nodes deployed on track sleepers, which are mainly responsible for collecting data such as track vibration, temperature and air pressure. The transmission layer utilizes LoRa communication technology to realize data transmission between nodes and uploads the data to the cloud server through the 4G module. The application layer stores, analyzes and visualizes the collected data to assist the operation and maintenance personnel in making decisions.In terms of design ideas, the system emphasizes low power consumption, modularity and scalability. By adopting low-power sensors and microcontrollers and combining them with reasonable power management strategies, the endurance of the system is extended. The modularized design makes the system easy to maintain and upgrade, and meets the demand for future function expansion. In addition, the system also considers the complexity of the deployment environment to ensure that the equipment has good waterproof and dustproof performance to adapt to different application scenarios, such as urban railways and mountainous railroads.

The system adopts “end-edge-cloud” three-level architecture mode:
1.	End (wireless sensor node): deployed in the key parts of the track, used to collect track vibration, temperature, air pressure and other raw state data, and sent through LoRa;
2.	Edge (gateway module): responsible for protocol conversion and data relay between LoRa network and 4G cellular network;
3.	Cloud (server platform): receives, stores and analyzes track data, provides visualization display, history record retrospective, threshold warning and other functions.

The system selects STM32F103C8T6 as the main control chip, which has enough computing power and rich peripheral interface resources, and is suitable for a variety of functional requirements such as LoRa communication control, sensor data acquisition and system power management. Software development is carried out through STM32CubeIDE, and the abstraction control of hardware is realized using HAL library.

Adopting LM401 LoRa module, based on Semtech SX1278 RF chip, it supports 433 MHz frequency band, the maximum communication distance is up to 5 kilometers (line-of-sight), and supports star network structure. Connected with STM32 MCU through SPI interface, the module power consumption is well controlled, the idle listening power consumption is about 15 mA, and the transmitting power consumption does not exceed 120 mA.ADXL362 Acceleration Sensor: supports 3-axis measurement, built-in FIFO, supports wake-up mode and motion detection mode, with power consumption as low as 2 µA. connected to MCU via SPI1 interface, it is used for detecting track micro-oscillation and shock signals, with sampling frequency up to 400 Hz.Temperature sensor (LMT85): A high-accuracy, low-power analog temperature sensor for temperature measurement from -50°C to 150°C has been introduced. Manufactured in a CMOS process, the sensor's output voltage is linearly inversely proportional to temperature, making it easy to read temperature through the ADC interface of a microcontroller (MCU).
Barometric Pressure Sensor (RSCM17100KP101): Suitable for positive pressure measurement with a range of 0 - 100 kPa. The module integrates a bridge sensor and a signal conditioning circuit with differential amplification, automatic calibration and temperature compensation, and is capable of directly outputting an analog voltage signal (e.g., 0.5 - 4.5 V), which makes it easy for the microcontroller (MCU) to read the pressure value through the ADC interface.Uses a rechargeable lithium battery with a voltage regulator module with a capacity of 2000mAh and a voltage of 5V.

The system adopts LoRa “star” networking mode, where multiple sensor nodes send data to the same gateway. To ensure reliable data transmission and anti-interference, the system adopts the following communication strategy:
(1)	Each node sets a unique device ID, the wireless sensor node ID from 6 to 9, and the gateway module ID is 10;
(2)	The data packet contains check digit, time stamp and sensor status identification, which is convenient for decoding and abnormal identification at the gateway;
(3)	Communication parameters are set as follows: spreading factor SF = 12, bandwidth BW=500kHz, coding rate CR = 4/5, LORA leading code length of 8 bits, receive timeout bits of 5 bits, variable packet length, and the maximum air rate is about 1.17kbps.
Other technical parameters are: frequency range 433 MHz, transmit power +22 dBm, receive sensitivity down to -148 dBm, SPI interface.

\begin{figure}[h]
	\centering
		\includegraphics[scale=1]{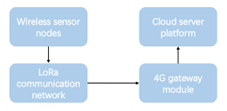}
	  \caption{Overall System Architecture}\label{FIG:1}
\end{figure}

\begin{figure}[h]
	\centering
		\includegraphics[scale=1]{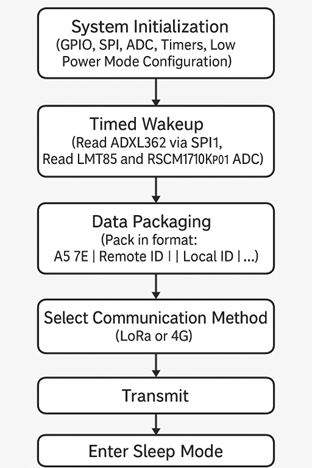}
	  \caption{System Main Cycle Block Diagram}\label{FIG:2}
\end{figure}

\section{Software system design and communication protocols}

The software architecture of this system is based on the STM32F103C8T6 as the core, which coordinates the data acquisition, processing and remote uploading of the ADXL362 tri-axial acceleration sensor, the LMT85 temperature sensor and the RSCM17100KP101 barometric pressure sensor through the three interfaces, namely, the SPI1 channel, the ADC channel and the UART. After system startup, the system first performs peripheral initialization SPI1, the ADC channels are configured sequentially to collect the analog voltages of LMT85 and RSCM17100KP101; UART1 is ready at 115200 bps baud rate, and the receiving end UART1 is directly connected to the 4G module. After the initialization is completed, STM32 automatically recovers and reads the three-axis acceleration, temperature and barometric pressure data in turn, and enters the data encapsulation and communication process. After completing the data sending, the system enters a low-power mode lower than the sleep mode and waits for a set time before restarting the execution of the main program.

In order to realize lightweight and unified data transmission between LoRa and 4G networks, the system adopts a fixed 20-byte length data frame format. Each frame has a two-byte header (0xA5, 0x7E) as the synchronization flag, followed by a one-byte remote device ID (fixed at 0x10) and a one-byte local node ID, and then a one-byte reserved field for future protocol upgrades. The X, Y, and Z acceleration data (4 bytes per axis) are then written sequentially, occupying 12 bytes in total, followed by the temperature and barometric pressure values, each of which is written as a 16-bit integer, occupying 2 bytes each. No checksum is used in the entire frame to minimize transmission latency and reduce processing overhead. The transmitting system tries to send the original 20-byte frame through the LM401 LoRa module, and when the receiving end receives the signal, the data is interpreted and converted to JSON format by the module internally, and then the same frame data is directly passed to the 4G module through UART1, and then uploaded to the server in the cloud.

After the system is powered on or reset, the STM32F103C8T6 main controller first initializes the peripherals such as system clock, GPIO, SPI, ADC and UART. The system master frequency is set to 72 MHz to guarantee the real-time data acquisition and communication process.The SPI1 interface is configured as the master mode, working in SPI mode 0 (CPOL = 0, CPHA = 0), with low clock polarity and rising clock sample edge, in order to adapt to the communication requirements of the ADXL362 acceleration sensor. The SPI clock frequency is set as the master frequency divided by 8, i.e., 9 MHz, which is in line with the maximum SPI clock frequency limitation supported by the ADXL362 (10 MHz or less). below) supported by the ADXL362.The ADXL362 initialization process is completed via SPI, including reset, setting to measurement mode, closing the FIFO buffer, configuring the bandwidth and output data rate (default ODR = 100 Hz), and initiating autosampling. In the master control, use the query method to continuously read the acceleration data of X, Y, and Z axes from the ADXL362, and the raw data obtained is a signed 16-bit integer in LSB, which is converted to get the physical acceleration value (in mg). The interrupt function of ADXL362 is not enabled in this system, and all sampling actions are initiated by the master control.

\section{System Testing and Performance Evaluation}

\begin{figure}[h]
	\centering
		\includegraphics[scale=0.4]{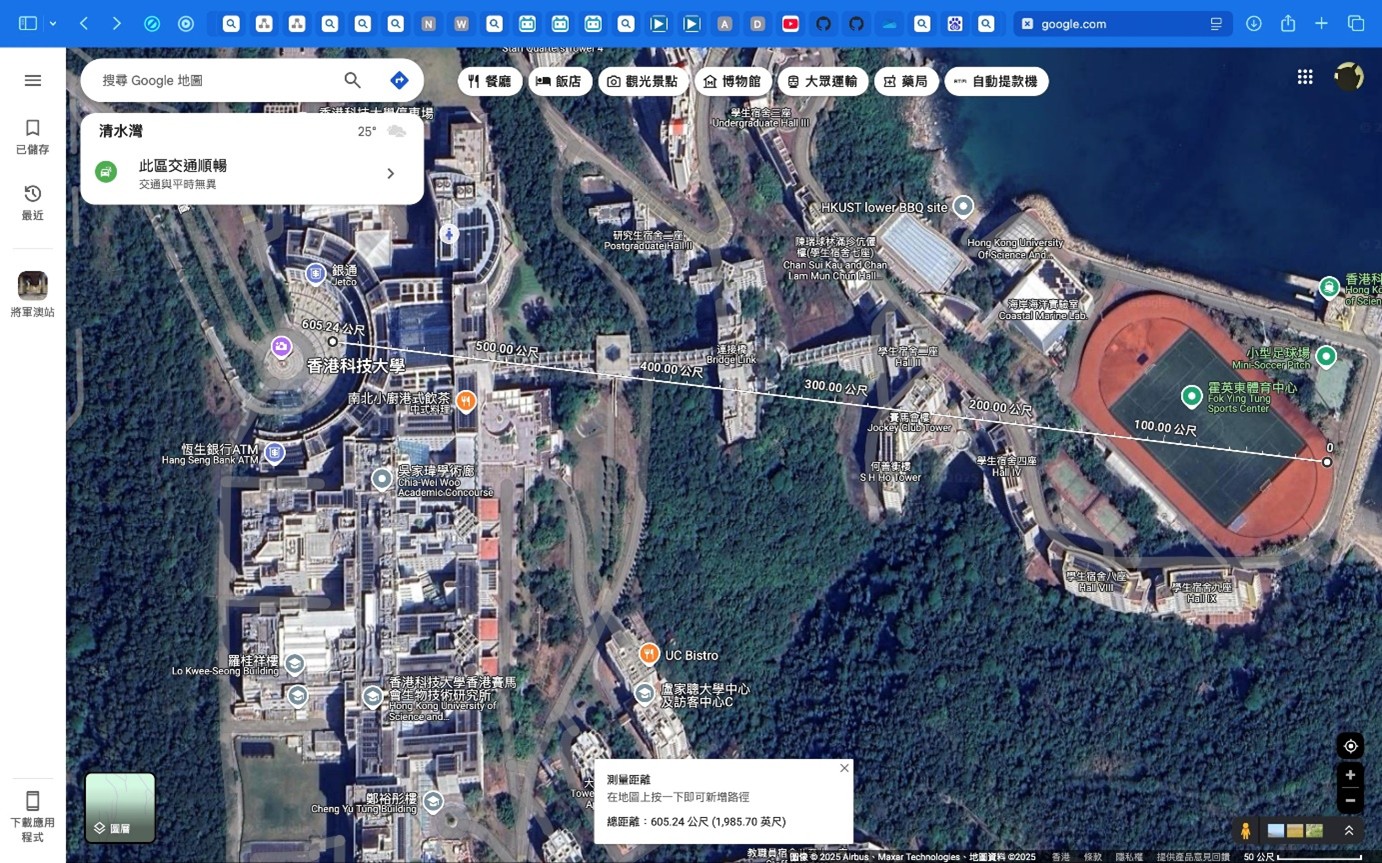}
	  \caption{Communication Distance Test Scenario Explanation:a}\label{FIG:3}
\end{figure}

\begin{figure}[h]
	\centering
		\includegraphics[scale=0.5]{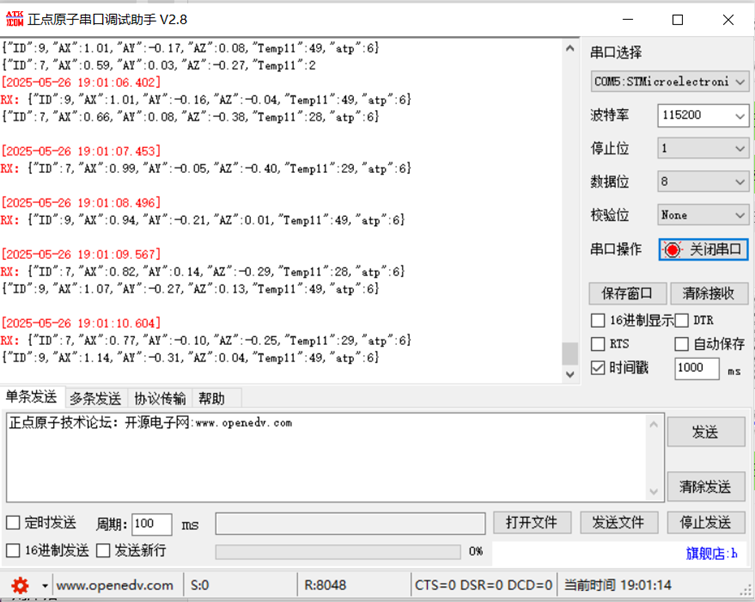}
	  \caption{Communication Distance Test Scenario Explanation:b}\label{FIG:4}
\end{figure}

In order to verify the functional integrity and communication performance of the designed wireless monitoring system for railroad tracks in real scenarios, this paper carries out indoor functional tests and outdoor communication evaluation of the prototype system. The tests cover the correctness of sensor data acquisition, the stability of SPI and serial communication, and the actual coverage ability of LoRa wireless communication. The test results provide a basis for the subsequent deployment and application of the system in a real track environment.

In the function verification stage, the system first builds a test platform in the laboratory environment, and completes the initialization of the sensor module and the data acquisition process through the STM32 master control chip. the ADXL362 acceleration sensor adopts the SPI interface to communicate with the master control, and configures the registers, starts the measurement and reads the three-axis data through the SPI1 bus in the actual operation. After the system is powered on, the first step is to verify the communication status of the SPI bus and determine whether the sensor responds normally by reading the device ID (0xAD) register of the ADXL362. The test results show that the SPI communication link is stable and the readout value is accurate, indicating that the initialization and operation logic of SPI communication is correct and can support continuous data interaction.

In addition, the system periodically starts the ADC conversion process to collect the analog voltage signals from the LMT85 temperature sensor and the RSCM17100KP101 barometric pressure sensor respectively. The sampled data is converted and packaged inside the MCU and encapsulated in a predefined communication protocol format. Finally, the master control sends the packaged data to the 4G communication module through the serial port (baud rate 115200) to complete the remote data upload. During the test, the debugging information of the serial port completely reflects the acquisition and uploading process of each cycle, and the system has a complete functional link and works stably as a whole.

In order to verify the communication coverage ability of LoRa module in outdoor track-like environment, this paper adopts the field step-by-step testing method to evaluate the communication distance of the prototype system in the open area in four rounds. During the test, the system works in LoRa communication transmit power 22 dBm mode, deployed by one node fixedly and carried by another node mobile, and each round is gradually pulled farther from the near distance, observing the data reception status and recording the farthest effective communication distance.

The test is conducted four times, and the results show that under the condition of no obstruction interference, the LoRa module can realize the farthest communication distance of X meters. Within this distance, the data packets can be received completely with acceptable delay, which indicates that the SX127x series LoRa modules have the ability to be deployed in railroad track monitoring scenarios. Considering that the rail environment is relatively linear and open, this communication performance provides theoretical support for the spacing design of subsequent node deployment.

In summary, all functional modules of the system have completed preliminary verification under the experimental conditions in this phase. Sensor data can be collected stably, SPI communication is accurate, data encapsulation logic is clear, 4G module serial port connection is reliable, and LoRa communication link has practical coverage capability, which lays a foundation for subsequent environmental adaptability test, long-term deployment experiment and system expansion and upgrading.

\section{Summary and Prospect}

In this study, a railroad track wireless monitoring system based on the fusion of LoRa communication technology and multiple sensors is designed and realized, aiming to meet the urgent needs of automation, high frequency, low cost and remote data acquisition in the current track health monitoring process. The system takes STM32 microcontroller as the core, integrates ADXL362 tri-axial acceleration sensor, LMT85 temperature sensor and RSCM17100KP101 barometric pressure sensor, and realizes remote uploading of on-site data with 4G module, and adopts low-power consumption working mode and timed data acquisition mechanism, and the LoRa wireless module provides long-distance data transmission channel, which makes the system more convenient without relying on the existing power and power supply of the track. LoRa wireless module provides a long-distance data transmission channel, which makes the system capable of independent operation without relying on the existing power and communication facilities of the railroad.

In summary, this study has completed the design, realization and validation of the prototype system with the practical function realization as the core, and has achieved preliminary results in key technology points such as track condition monitoring, low-power communication and remote data uploading. In the future, we will continue to optimize and improve the practicality, intelligence and scalability of the system, so as to promote the development of the system in the direction of practical deployment and industrial application.










\bibliographystyle{cas-model2-names}

\bibliography{cas-refs}



\end{document}